\documentclass[11pt]{article}

\usepackage[utf8]{inputenc}
\usepackage[english]{babel}
\usepackage{amsthm}

\usepackage{graphicx}
\usepackage{amsmath}
\usepackage{amssymb}
\usepackage{lscape}
\usepackage{bm}

\makeatletter
\@addtoreset{equation}{section}

\makeatother

\begin{document}

\title{
\vbox{
\baselineskip 14pt
\hfill \hbox{\normalsize KEK-TH-2219
}} \vskip 1cm
\bf \Large  
More on Effective Potentials for Revolving D-Branes
\vskip .5cm
}

\author{
Satoshi Iso$^{a,b}$  \thanks{E-mail: \tt iso(at)post.kek.jp}, 
Noriaki Kitazawa$^{c}$ \thanks{E-mail: \tt noriaki.kitazawa(at)tmu.ac.jp},
Hikaru Ohta$^{a}$ \thanks{E-mail: \tt hohta(at)post.kek.jp}, 
Takao Suyama$^{a}$ \thanks{E-mail: \tt tsuyama(at)post.kek.jp}
\bigskip\\
\it \normalsize
$^a$ Theory Center, High Energy Accelerator Research Organization (KEK), \\
\it  \normalsize 
$^b$Graduate University for Advanced Studies (SOKENDAI),\\
\it \normalsize Tsukuba, Ibaraki 305-0801, Japan \\
\\
\it  \normalsize 
$^c$Department of Physics, Tokyo Metropolitan University,\\
\it  \normalsize Hachioji, Tokyo 192-0397, Japan \\
\smallskip
}
\date{\today}

\maketitle

\abstract{\normalsize
We continue to investigate the effective potential between a pair of D$p$-branes revolving around each other
by using the technique of {\it partial modular transformation} developed
in our previous work.
We determine the shape of the potential for general $p$ for a wide range of regions interpolating 
 smaller and larger distances than the string scale $l_s$. 
We also discuss the backreaction of the D-brane system to the space-time metric and the validity of our calculations.
}

\newpage


\newpage

\section{Introduction}

In this paper, we continue to investigate a system of D-branes which are revolving around each other discussed in \cite{Iso:2019gbd}. 
Our motivation in these studies
is to apply such a D-brane system to stringy phenomenological models of electroweak symmetry breaking \cite{Blumenhagen:2006ci,Ibanez:2012zz,Baumann:2014nda},
in which  a small energy scale, compared to the UV stringy scale, may emerge due to a dynamical motion of D-branes. 
In particular, 
we hope to realize the weak scale 
as the radius $r$ of the circular orbit determined by the angular velocity of  the revolving D-brane system\cite{Iso:2015mva,Iso:2018sgx}. 

For this purpose, in \cite{Iso:2019gbd} we developed an efficient method to calculate
the effective potential between the revolving D$p$-branes in a wide region of their distance, which we call {\it partial modular transformation}.
By employing this method, 
we obtained an expression of the effective potential for general $p$ as a sum of complicated integrals, 
which will be recalled in section \ref{potential}. 
In \cite{Iso:2019gbd}, only the $p=3$ case  was analyzed and the shape of the effective potential was explicitly obtained
 since this seems to be the most interesting from a phenomenological point of view. 

In this paper, we extend the analysis of the $p=3$ case  to more general D$p$-branes.
Since the effective potential for large radius $r$ can be obtained rather easily by using the Type II supergravity (SUGRA), 
we focus on the region of $r$ smaller than the string scale $l_s := \sqrt{2 \pi \alpha'}$, where we need to use super Yang-Mills (SYM) theory as an effective theory. 
In particular, for $r \sim l_s$, we need to sum both contributions from supergravity and super Yang-Mills theory.
If we simply sum these two, it causes a problem of double counting.  
The method of the partial modular transformation can avoid this problem, and we can obtain the 
effective potential in a good approximation interpolating from large $r>l_s$ to small $r<l_s$. 

We find that the shape of the potential for general D$p$-branes is qualitatively similar to the one for $p=3$, 
that is, an attractive force dominating the repulsive centrifugal force, appears for a certain situation. 
More specifically, the behaviors of the potentials for $p=0,1$ are different from the others. 
The result for $p=0$ indicates  the presence of non-BPS bound states of D0-branes discussed in \cite{Danielsson:1996uw, Kabat:1996cu}. 
We also discuss the validity of our calculation, in particular the effect of the backreaction of the whole D-brane system to the background space-time metric. 
The behavior of the potential at  $r \ll l_{s}$  is also discussed. 

The paper is organized as follows. 
In section \ref{partial}, we recall the results of our previous paper \cite{Iso:2019gbd}. 
The integral expressions of the effective potential are explicitly evaluated in section \ref{potential}. 
The shape of the effective potential, including the centrifugal potential, is examined in section \ref{shape}.
The behavior of the effective potential for extremely small $r$ is analyzed in section \ref{conclusion}. 
Some technical details are summarized in Appendices. 

\section{Partial modular transformation}
\label{partial}

In this section, we briefly review our previous work \cite{Iso:2019gbd}. 
First, we recall the method for calculating the effective potential between revolving D-branes. 

Interactions between dynamical D-branes are mediated by exchange of closed strings. 
At the leading order in $g_s$, the effective potential can be also calculated by the one-loop amplitude of an open string stretched 
between the D-branes. 
Such calculations for static D-branes \cite{Polchinski:1998rq, Polchinski:1998rr} are well-known. 
Schematically, it is given as 
\begin{equation}
V(R)\ =\ \int_0^\infty\frac{dt}{t}Z_{\rm open}(t,R), 
   \label{open 1-loop}
\end{equation}
where $R$ is the distance between the D-branes, which may vary with time in general. 
For D-branes in motion, calculations are more difficult to perform;
when they move at a constant relative velocity,  weak attractive force is obtained \cite{Bachas:1995kx, Lifschytz:1996iq}. 
For general motions, only perturbative calculations are performed \cite{Hirano:1996pf,Kazama:1997bc}. 
When D-branes are revolving around each other, we obtained attractive force by using a perturbative method \cite{Iso:2018cwb}.

For performing the calculation in a more efficient way, we developed a new technique in \cite{Iso:2019gbd}.
We first rewrite this into the following sum of open and closed string contributions (see also \cite{Douglas:1996yp});
\begin{equation}
V(R)\ =\ \int_1^\infty\frac{dt}{t}Z_{\rm open}(t,R)+\int_1^\infty ds\,Z_{\rm closed}(s,R), 
\end{equation}
which is obtained from (\ref{open 1-loop}) by dividing the integration region into $[0,1)$ and $[1,\infty)$, and performing the modular transformation $s=t^{-1}$ only for the integration region over $[0,1)$. 
We call this operation the {\it partial modular transformation}. 
Note that $Z_{\rm closed}(s,R)$ gives the amplitude for exchanging one closed string between the D-branes. 

In this expression, we can safely ignore contributions from all heavy states in $Z_{\rm open}(t,R)$ 
and $Z_{\rm closed}(s,R)$ as far as 
a mass gap exits between lightest (massless) states and heavy massive states.
It does hold irrespective of the value of $R$ from $R \gg l_s$ to $R \ll l_s$.
Calculating the contributions of light states in both open and closed string sectors turns out to give
 an accurate approximation $\tilde{V}(R)$ for the effective potential $V(R)$. 
 In \cite{Iso:2019gbd}, we see that its accuracy to the original potential $V(R)$ is  just a few percents for an arbitrary value of $R$.

By construction, $\tilde{V}(R)$ is schematically given by the form 
\begin{equation}
\tilde{V}(R)\ =\ \mbox{(1-loop in SYM)}+\mbox{(tree-level in SUGRA)},
\end{equation}
where each string contribution is replaced by its lowest modes.
An important point here is that the Schwinger parameters $t, s$ have cut-offs which allow us to avoid a double-counting of contributions to $\tilde{V}(R)$. 
Note that the calculation of the right-hand side is possible even when we do not know how to quantize an open string stretched between the dynamical D-branes. 
It is also interesting to notice that the SUGRA contribution can be interpreted as threshold corrections to the SYM contribution since the former represents effects coming from heavy open string states. 

\vspace{5mm}

In \cite{Iso:2019gbd}, we applied the above method to a pair of D$p$-branes revolving around each other. 
Let $r$ and $\omega$ be the radius of the circular orbit and the angular velocity of the D$p$-branes, respectively, so that 
we have $R=2r$ in this case. 
We obtained an integral formula which gives a good approximation $\tilde{V}_p(2r)$ for the effective potential between the revolving D$p$-branes. 
For a systematic evaluation of the potential, we divide $\tilde{V}_p(2r)$ into 
 the following three contributions:
bosonic fields in open strings $\tilde{V}_{B,p}(2r)$, fermionic fields in open strings $\tilde{V}_{F,p}(2r)$ 
and closed string contributions $\tilde{V}_{c,p}(2r)$,
\begin{equation}
\tilde{V}_p(2r)\ =\ \tilde{V}_{B,p}(2r)+\tilde{V}_{F,p}(2r)+\tilde{V}_{c,p}(2r). 
\end{equation}
The bosonic fields, including ghosts,  in SYM give the contributions,
\begin{eqnarray}
\tilde{V}_{B,p}(2r) 
&=& -\int_{\Lambda^{-2}}^\infty\frac{dt}{t}\int\frac{d^{p+1}k}{(2\pi)^{p+1}}\,e^{-t(k^2+4r^2)} \nonumber \\
& & \times \left[ 6+2e^{-t(\omega^2-\frac{8(r\omega)^2}{k^2+4r^2})}\cosh\left( t\sqrt{4\omega^2k_\tau^2+\left( \frac{8(r\omega)^2}{k^2+4r^2} \right)^2} \right) \right], 
   \label{SYM-boson}
\end{eqnarray}
where $\Lambda:=1/\sqrt{2\pi\alpha'}=1/l_s$ and $k_\tau$ is the time component of Euclidean $(p+1)$-vector $k$. 
The fermionic fields in SYM give
\begin{equation}
\tilde{V}_{F,p}(2r)\ =\ 8\int_{\Lambda^{-2}}^\infty\frac{dt}{t}\int\frac{d^{p+1}k}{(2\pi)^{p+1}}\,e^{-t(k^2+4r^2)}e^{-\frac14t\omega^2}\cosh\left( t\sqrt{\omega^2k_\tau^2+4(r\omega)^2} \right), 
   \label{SYM-fermion}
\end{equation}
and the contributions from the bosonic fields in Type II SUGRA are given by
\begin{eqnarray}
\tilde{V}_{c,p}(2r) 
&=& -\kappa_{10}^2T_p^2(4\pi)^{-\frac{10-p}{2}}\frac{(r\omega)^2}{1+(r\omega)^2}\int_{\tilde{\Lambda}^{-2}}^\infty ds\,s^{-\frac{10-p}{2}} \nonumber \\
& & \times\int_{-\infty}^{+\infty}d\zeta\,\exp\left[ -\frac1{4s}\left( \zeta^2+2r^2(1+\cos\omega\zeta) \right) \right](1+\cos\omega\zeta)^2, 
   \label{SUGRA}
\end{eqnarray}
where $\tilde{\Lambda}:=2/\sqrt{2\pi\alpha'}$, and $\kappa_{10}^2T_p^2=\pi(2\pi l_s^2)^{3-p}$. 
These integrals have various subtle behaviors when we expand them as a power series of either $r/\omega$ or $\omega/r$. 
In the following, we explicitly evaluate them to analyze the shape of $\tilde{V}_p(2r)$ for general $p$. 
The case $p=3$ was discussed in \cite{Iso:2019gbd}.  

Note that the above expressions are derived in the Euclidean signature. 
When we discuss the dynamics of D$p$-branes in this potential, we will perform a suitable analytic continuation of $\omega$. 
Note also that $r$ in $\tilde{V}_{B,p}(2r)$ and $\tilde{V}_{F,p}(2r)$ has the mass dimension $+1$, while $r$ in $\tilde{V}_{c,p}(2r)$ has $-1$. 
Therefore, we will replace $r$ with $r/l_s^2$ for the former at the end of calculations. 
We believe that it will not make any confusion.

\section{Effective potential for small $\omega$}
\label{potential}

In this section,  
we will evaluate the integrals in eqs. (\ref{SYM-boson}), (\ref{SYM-fermion}) and (\ref{SUGRA})
to investigate the shape of the approximate potential $\tilde{V}(2r)$ for each $p$.
In particular, we are interested in the situation where $\omega$ is small and the motion is nonrelativistic $r \omega \ll 1$. 
In the following, we expand the integrands in
eqs. (\ref{SYM-boson}), (\ref{SYM-fermion}) and (\ref{SUGRA}) with respect to $\omega$, and then perform the integrals term by term. 

\subsection{Contributions from SYM}
\label{sec:SYM}

By rescaling the integration variables, the integrals eqs. (\ref{SYM-boson}) and (\ref{SYM-fermion}) are functions of $\omega/r$, up to some overall factors, and we can expand them in terms of $\omega/r$. 
Therefore, the following calculations are valid as long as $\omega\ll r$ is satisfied.

First, let us consider $\tilde{V}_{B,p}(2r)$. 
By expanding the integrands with respect to $\omega/r$, the $t$-integration can be performed
and we have
\begin{align} 
\tilde V_{B,p}(2r)=&\ c(r)+\int\frac{d^{p+1}k}{(2\pi)^{p+1}}e^{-(k^2+4r^2)/\Lambda^2}\left[ \omega^2\left( \frac{2\Lambda^2-4k_\tau^2}{\Lambda^2(k^2+4r^2)}-\frac{16r^2+4k_\tau^2}{(k^2+4r^2)^2} \right) \right. \notag \\[2mm] 
&+\omega^4\left( -\frac{3\Lambda^4-12\Lambda^2k_\tau^2+4k_\tau^4}{3\Lambda^6(k^2+4r^2)} +\frac{16\Lambda^2r^2-32k_\tau^2r^2-\Lambda^4+8k_\tau^2\Lambda^2-4k_\tau^4}{\Lambda^4(k^2+4r^2)^2} \right. \notag \\[2mm] 
& \left. \left. -\frac{128r^4-16\Lambda^2r^2+64k_\tau^2r^2-8k_\tau^2\Lambda^2+8k_\tau^4}{\Lambda^2(k^2+4r^2)^3}-\frac{128r^4+64k_\tau^2r^2+8k_\tau^4}{(k^2+4r^2)^4} \right) \right] \notag \\
&+{\cal O}(\omega^6), 
\end{align}
where $c(r)$ is an ${\cal O}(\omega^0)$ contribution. 
This term is  canceled by an ${\cal O}(\omega^0)$ term in $\tilde{V}_{F,p}(2r)$ since the system with $\omega=0$ is a BPS configuration. 
In the same manner, we obtain 
\begin{align}
\tilde V_{F,p}(2r)=&\ -c(r)+\int\frac{d^{p+1}k}{(2\pi)^{p+1}}e^{-(k^2+4r^2)/\Lambda^2}\left[ \omega^2\left( \frac{16r^2-2\Lambda^2+4k_\tau^2}{\Lambda^2(k^2+4r^2)}+\frac{16r^2+4k_\tau^2}{(k^2+4r^2)^2} \right) \right. \notag \\[2mm] 
&\left. +\omega^4\left( \frac{64r^4-48\Lambda^2r^2+32k_\tau^2r^2+3\Lambda^4-12k_\tau^2\Lambda^2+4k_\tau^4}{12\Lambda^6(k^2+4r^2)} \right. \right. \notag \\[2mm] 
& +\frac{64r^4-32\Lambda^2r^2+32k_\tau^2r^2+\Lambda^4-8k_\tau^2\Lambda^2+4k_\tau^4}{4\Lambda^4(k^2+4r^2)^2} \notag \\[2mm] 
& \left. \left. +\frac{32r^4-8\Lambda^2r^2+16k_\tau^2r^2-2k_\tau^2\Lambda^2+2k_\tau^4}{\Lambda^2(k^2+4r^2)^3}+\frac{32r^4+16k_\tau^2r^2+2k_\tau^4}{(k^2+4r^2)^4} \right) \right] \notag \\
& +{\cal O}(\omega^6). 
\end{align}
Summing them, we obtain the following SYM part of the potential,
\begin{align}
\tilde V_{o,p}(2r):=
&\ \tilde{V}_{B,p}(2r)+\tilde{V}_{F,p}(2r) \notag \\[2mm]
=&\ \int\frac{d^{p+1}k}{(2\pi)^{p+1}}e^{-(k^2+4r^2)/\Lambda^2}\left[ \omega^2\,\frac{16r^2}{\Lambda^2(k^2+4r^2)}\right. \notag \\[2mm] 
& +\omega^4\left( \frac{-9\Lambda^4-48r^2\Lambda^2+64r^4+(36\Lambda^2+32r^2)k_\tau^2-12k_\tau^4}{12\Lambda^6(k^2+4r^2)} \right. \notag \\[2mm]
&+\frac{-3\Lambda^4+32\Lambda^2r^2+64r^4+(24\Lambda^2-96r^2)k_\tau^2-12k_\tau^4}{4\Lambda^4(k^2+4r^2)^2}\notag \\[2mm] 
& \left. \left. +\frac{8\Lambda^2r^2-96r^4+(6\Lambda^2-48r^2)k_\tau^2-6k_\tau^4}{\Lambda^2(k^2+4r^2)^3}+\frac{-96r^4-48k_\tau^2r^2-6k_\tau^4}{(k^2+4r^2)^4} \right) \right] \notag \\
& +{\cal O}(\omega^6).
\label{V_{O,p}} 
\end{align}

To perform the $k$-integration, it is convenient to use the following formula,
\begin{align}
&\int\frac{d^{p+1}k}{(2\pi)^{p+1}}f(k^2)k_\tau^{2m}
=\frac1\pi(4\pi)^{-\frac{p}2}\frac{\Gamma(m+\frac12)}{\Gamma(m+\frac{p+1}2)}
\int_0^\infty dk\, f(k^2)k^{2m+p},
\label{k_int}
\end{align}
where $k_\tau$ is a Euclidean-time component of the momentum $k$. 
This is obtained by employing the polar coordinates for $k$ and performing the integration for the angular coordinates. 
Then, we obtain 
\begin{eqnarray}
\tilde{V}_{o,p}(2r) 
&=& \frac1\pi(4\pi)^{-\frac{p}2}\frac{\sqrt{\pi}}{\Gamma(\frac{p+1}2)}\int_0^\infty dk\,e^{-(k^2+4r^2)/\Lambda^2} \nonumber \\
& & \times\left[ \omega^2\frac{16r^2}{\Lambda^2}\frac{k^p}{k^2+4r^2}+\omega^4\sum_{m=-1}^4\frac{c_m(2r/\Lambda)}{\Lambda^{4-2m}}\frac{k^p}{(k^2+4r^2)^m} \right]+{\cal O}(\omega^6), 
\end{eqnarray}
where the functions $c_m(x)$ are defined as 
\begin{eqnarray}
c_{-1}(x) 
&:=& -g_4, \\ [2mm] 
c_0(x) 
&:=& \frac{6g_4+2g_2}{3}x^2+3(g_2-g_4), \\ [2mm] 
c_1(x) 
&:=& -\frac{3g_4+2g_2-1}{3}x^4+(6g_4-9g_2-1)x^2-\frac{24g_4-24g_2+3}{4}, \\ [2mm] 
c_2(x) 
&:=& (-3g_4+6g_2+1)x^4+(12g_4-18g_2+2)x^2-\frac{24g_4-24g_2+3}{4}, \\ [2mm] 
c_3(x) 
&:=& -(6g_4-12g_2+6)x^4+(12g_4-18g_2+2)x^2, \\ [2mm]
c_4(x) 
&:=& -(6g_4-12g_2+6)x^4, 
\end{eqnarray}
and $g_2:=1/(p+1)$, $g_4:=3/(p+1)(p+3)$. 
By defining the following functions 
\begin{equation}
f_{p,m}(x)\ :=\ x^{1+p-2m}\int_0^\infty dk\,e^{-x^2(k^2+1)}\frac{k^p}{(k^2+1)^m}, 
\end{equation}
$\tilde{V}_{o,p}(2r)$ can be rewritten as 
\begin{eqnarray}
\tilde{V}_{o,p}(2r) 
&=& \frac1\pi(4\pi)^{-\frac{p}2}\frac{\sqrt{\pi}}{\Gamma(\frac{p+1}2)}\left[ 4\omega^2\Lambda^{p-1}\left( \frac{2r}{\Lambda} \right)^2f_{p,1}\left( \frac{2r}\Lambda \right)+\omega^4\Lambda^{p-3}\sum_{m=-1}^4c_m\left( \frac{2r}{\Lambda} \right)f_{p,m}\left( \frac{2r}{\Lambda} \right) \right] \nonumber \\
& & +{\cal O}(\omega^6). 
\end{eqnarray}
Note that the functions $f_{p,m}(x)$  satisfy the following recurrence relation,
\begin{equation}
f_{p,m}(x)\ =\ f_{p-2,m-1}(x)-x^2f_{p-2,m}(x). 
\end{equation}
Thus, it is enough to determine $f_{0,m}(x)$ or $f_{1,m}(x)$ depending on whether $p$ is even or odd. 
The details of these functions are given in Appendix \ref{app-SYM}. 

\vspace{5mm}

We apply the formulas given in Appendix \ref{app-SYM}, and replace $r$ by $r/l_s^2$ such that $r$ has the dimension of length. 
Returning to the Lorentzian signature by performing the analytic continuation of $\omega$, the potential is obtained as follows. 
For even $p$, 
\begin{align}
\tilde V_{o,0}(2r)
=&-\frac1{\sqrt{\pi}}\left[4\sqrt{\pi} \omega^2 r \left(1-\textrm{Erf}\left(\frac{2r}{l_s}\right)\right)\right. \notag  \\
&\left. \hspace{5mm}-\omega^4\left\{\left(-\frac{15\, l_s^5}{32 r^2}-\frac{5\, l_s^3}{4}+\frac{34\, l_sr^2}{3}\right)e^{-4r^2/l_s^2}-\sqrt{\pi}\left(\frac{15\, l_s^6}{128r^3}+\frac{64r^3}{3}\right)\left(1-\textrm{Erf}\left(\frac{2r}{l_s}\right)\right)\right\}\right] \notag \\
&\hspace{5mm}+{\cal O}(\omega^6),
\end{align}
\begin{align}
\tilde V_{o,2}(2r)
=&-\frac1{\pi^{3/2}}\left[
\omega^2\left\{\frac{4r^2}{l_s^3}e^{-4r^2/l_s^2}-\frac{8\sqrt{\pi}r^3}{l_s^4}\left(1-\textrm{Erf}\left(\frac{2r}{l_s}\right)\right)\right\} \right. \notag \\[2mm]
&\left. \hspace{1cm}
-\omega^4 
\left\{\left(-\frac{3\, l_s}{16}-\frac{37r^2}{18\, l_s}-\frac{128r^4}{9\, l_s^3}\right)e^{-4r^2/l_s^2}\right.\right. \notag \\[2mm]
&\left.\left. \hspace{2cm}
+\sqrt{\pi}\left(-\frac{3\,l_s^2}{64r}+\frac{8r^3}{\, l_s^2}+\frac{256r^5}{9\, l_s^4}\right)\left(1-\textrm{Erf}\left(\frac{2r}{l_s}\right)\right) \right\}\right]
+{\cal O}(\omega^6),
\end{align}
\begin{align}
\tilde V_{o,4}(2r)=&
-\frac1{\pi^{5/2}}\left[
\omega^2 \, \frac{r^2}{l_s^5}
\left\{\frac13\left(1-\frac{8r^2}{l_s^2}\right)e^{-4r^2/l_s^2}+\frac{16\sqrt{\pi}r^3}{3\,l_s^3}\left(1-\textrm{Erf}\left(\frac{2r}{l_s}\right)\right)\right\} \right. \notag \\[2mm]
&\left. \hspace{1cm}
-\omega^4 \, \frac{r}{l_s^2}
\left\{\left(-\frac{9r}{40\,l_s}+\frac{64r^3}{15l_s^3}+\frac{128r^5}{15l_s^5}\right)e^{-4r^2/l_s^2}\right.\right. \notag \\[2mm]
&\left.\left. \hspace{2cm}
-\sqrt{\pi}\left(\frac{1}{32}+\frac{32r^4}{3\,l_s^4}+\frac{256r^6}{15\,l_s^6}\right)\left(1-\textrm{Erf}\left(\frac{2r}{l_s}\right)\right) \right\}\right]
+{\cal O}(\omega^6),
\end{align}
\begin{align}
\tilde V_{o,6}(2r)=&
-\frac1{\pi^{7/2}}\left[
\omega^2 \, \frac{r^2}{l_s^7}
\left\{\left(\frac1{20}-\frac{2r^2}{15l_s^2}+\frac{16r^4}{15l_s^4}\right)e^{-4r^2/l_s^2}-\frac{32\sqrt{\pi}r^5}{15l_s^5}\left(1-\textrm{Erf}\left(\frac{2r}{l_s}\right)\right)\right\} \right. \notag \\[2mm]
&\left. \hspace{1cm}
-\frac{\omega^4}{l_s^3}
\left\{\left(-\frac{11r^2}{840l_s^2}+\frac{26r^4}{105l_s^4}-\frac{176r^6}{63l_s^6}-\frac{1024r^8}{315l_s^8}\right)e^{-4r^2/l_s^2}\right.\right. \notag \\[2mm]
&\left.\left. \hspace{2cm}
+\frac{\sqrt{\pi}r^3}{l_s^3}\left(-\frac1{16}+\frac{32r^4}{5l_s^4}+\frac{2048r^6}{315l_s^6}\right)\left(1-\textrm{Erf}\left(\frac{2r}{l_s}\right)\right) \right\}\right] \notag \\
&\hspace{5mm}+{\cal O}(\omega^6), 
\end{align}
%
and for odd $p$, 
\begin{align}
\tilde V_{o,1}(2r)
=&-\frac1\pi \left[\omega^2 \, \frac{4r^2}{l_s^2} E_1(4r^2/l_s^2) \right. \notag \\[2mm]
&\left. -\omega^4\left\{ \left( -\frac{l_s^4}{8r^2}-\frac{l_s^2}{2}+\frac{13r^2}{3} \right)e^{-4r^2/l_s^2}+\left(-2r^2-\frac{16r^4}{l_s^2} \right)E_1(4r^2/l_s^2)\right\} \right] \notag \\
&+{\cal O}(\omega^6),
\end{align}
\begin{align}
\tilde V_{o,3}(2r)
=&-\frac1 {\pi^2} \left[\omega^2\left\{ \frac{r^2}{l_s^4}e^{-4r^2/l_s^2}-\frac{4r^4}{l_s^6}E_1(4r^2/l_s^2) \right\} \right. \notag \\[2mm]
&\left. -\omega^4\left\{ \left( -\frac{1}{16}-\frac{7r^2}{12\, l_s^2}-\frac{10r^4}{3\, l_s^4} \right)e^{-4r^2/l_s^2}+\left( \frac{6r^4}{l_s^4}+\frac{40r^6}{3\, l_s^6} \right)E_1(4r^2/l_s^2) \right\} \right] \notag \\
&+{\cal O}(\omega^6),
\end{align}
\begin{align}
\tilde V_{o,5}(2r)
=&-\frac1{\pi^3} \left[\omega^2\left\{ \left(\frac{r^2}{8\, l_s^6}-\frac{r^4}{2\, l_s^8}\right)e^{-4r^2/l_s^2}+\frac{2r^6}{l_s^{10}}E_1(4r^2/l_s^2) \right\} \right. \notag \\[2mm]
&\left. -\omega^4\left\{ \left( -\frac{7r^2}{72\, l_s^4}+\frac{31r^4}{36\, l_s^6}+\frac{14r^6}{9\, l_s^8} \right)e^{-4r^2/l_s^2}+\left(-\frac{5r^6}{l_s^8}-\frac{56r^8}{9\, l_s^{10}} \right)E_1(4r^2/l_s^2) \right\} \right] \notag \\
&+{\cal O}(\omega^6),
\end{align}
\begin{align}
\tilde V_{o,7}(2r)
=&-\frac1{\pi^4}\left[\omega^2\left\{ \left(\frac{r^2}{48\, l_s^8}-\frac{r^4}{24\, l_s^{10}}+\frac{r^6}{6\, l_s^{12}}\right)e^{-4r^2/l_s^2}-\frac{2r^8}{3\, l_s^{14}}E_1(4r^2/l_s^2) \right\} \right. \notag \\[2mm]
&-\omega^4\left\{ \left( -\frac{r^2}{192\, l_s^6}+\frac{r^4}{12\, l_s^8}-\frac{11r^6}{24\, l_s^{10}}-\frac{r^8}{2\, l_s^{12}} \right)e^{-4r^2/l_s^2}\right. \notag \\[2mm]
&\hspace{1cm}+\left. \left. \left(-\frac{r^4}{16\, l_s^8}+\frac{7r^8}{3\, l_s^{12}}+\frac{2r^{10}}{ l_s^{14}} \right)E_1(4r^2/l_s^2) \right\} \right]
+{\cal O}(\omega^6). 
\end{align}

\subsection{Contributions from SUGRA}
\label{sec:SUGRA}

In the following, we expand the integrand of the formula (\ref{SUGRA}) with respect to $\omega$ in order to simplify the expression. 
It is obvious from the expression that the expansion starts with ${\cal O}(\omega^4)$ terms. 
Therefore, the contributions from SUGRA fields are sub-leading when $\omega$ is small. 

\begin{figure}[htpb]
\center
\includegraphics [scale=.8]
 {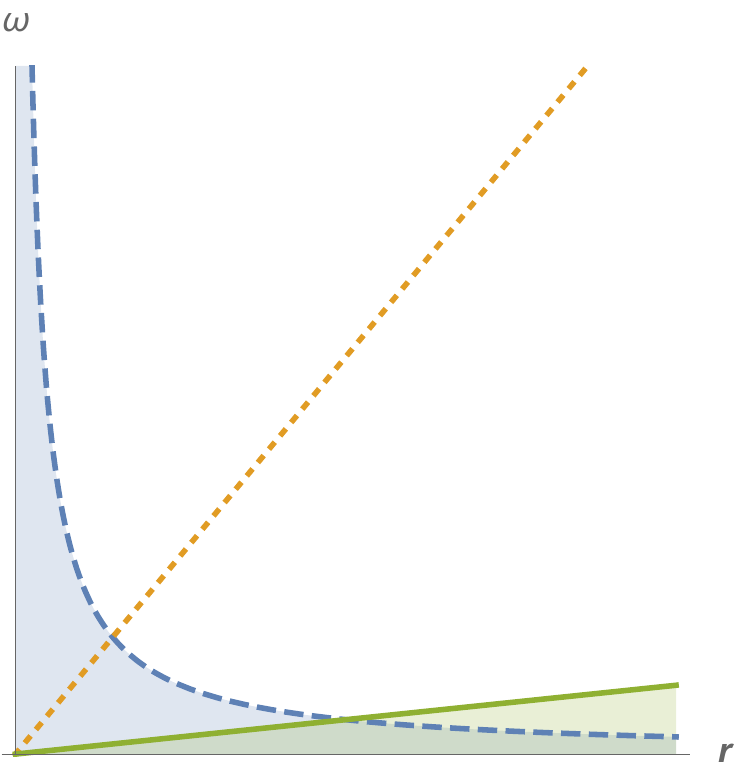}
\caption{Validity regions of SUGRA and SYM calculations. 
In blue region, the nonrelativistic condition, $\omega \ll 1/r$, is satisfied and our SUGRA
calculation is valid. 
On the other hand, in the green region, the SYM calculations in terms of $\omega l_s^2/r$ expansion is valid. 
We also depicted a line of $\omega=2r/\sqrt 3$, beyond which 
an instability may occur (see discussions in section \ref{conclusion}). 
Thus,  our analysis is valid in the overlapping region near the $\omega=0$ line. }
\label{Fig_omega-r}
\end{figure}

By a suitable rescaling of the integration variable, we find that the integrand depends only on the combination 
$v=r\omega$, up to some overall factor. 
Therefore, the following calculations are valid in a nonrelativistic region 
where $\omega\ll 1/r$ is satisfied. 
In Figure \ref{Fig_omega-r}, we draw this validity region as well as 
the validity region of the SYM calculations  $\omega\ll r/l_s^2$ discussed in 
section \ref{sec:SYM}. There are overlaps of these two regions, and hence
our analysis is valid there.

The leading order of  the expansion is given by
\begin{equation}
\tilde{V}_{c,p}(2r)\ =\ -\kappa_{10}^2T_p^2(4\pi)^{-(9-p)/2}\tilde{\Lambda}^{7-p}(r\omega)^4g_{(9-p)/2}\left( \frac{r\tilde{\Lambda}}{\sqrt{2}} \right)+{\cal O}(\omega^6), 
   \label{SUGRA expansion}
\end{equation}
where 
\begin{equation}
g_\alpha(x)\ :=\ \int_1^\infty ds\,s^{-\alpha}e^{-x^2/s}, 
\end{equation}
and we have performed the analytic continuation of $\omega$, which is in fact trivial. 
Note that the integral for $g_\alpha(x)$ is well-defined for $\alpha>1$, that is, $p<7$. 
They satisfy the following recurrence relations 
\begin{equation}
g_\alpha(x)\ =\ \frac{\alpha-2}{x^2}g_{\alpha-1}(x)-\frac1{x^2}e^{-x^2}. 
\end{equation}
Therefore, it is enough to determine $g_{3/2}(x)$ or $g_{2}(x)$ depending on whether $p$ is even or odd. 
The details of these functions are given in Appendix \ref{app-SUGRA}. 
According to this, the potential for even $p$ is obtained as 
\begin{align}
\tilde{V}_{c,0} (2r)
=&-\frac{\omega^4}{64\pi^{1/2}}\frac{l_s^5}{r^2}\notag \\[2mm]
&\times \left[-\left(30+20\left(\frac{2r}{l_s}\right)^2+32\left(\frac{2r}{l_s}\right)^4\right)e^{-4r^2/l_s^2}+15\sqrt{\pi}\left(\frac{2r}{l_s}\right)^{-1} \textrm{Erf}\left(\frac{2r}{l_s}\right)\right]+\mathcal{O}(\omega^6),\\[2mm]
\tilde{V}_{c,2} (2r)
=&-\frac{\omega^4l_s}{32\pi^{3/2}}
\left[-\left(6+4\left(\frac{2r}{l_s}\right)^2\right)e^{-4r^2/l_s^2}+3\sqrt{\pi}\left(\frac{2r}{l_s}\right)^{-1} \textrm{Erf}\left(\frac{2r}{l_s}\right)\right]+\mathcal{O}(\omega^6),\\[2mm]
\tilde{V}_{c,4} (2r)
=&-\frac{\omega^4}{16\pi^{5/2}}\frac{r^2}{l_s^3}
\left[-2e^{-4r^2/l_s^2}+\sqrt{\pi}\left(\frac{2r}{l_s}\right)^{-1} \textrm{Erf}\left(\frac{2r}{l_s}\right)\right]+\mathcal{O}(\omega^6),
\end{align}
\begin{align}
\tilde{V}_{c,6} (2r)
=&-\frac{\omega^4}{8\pi^{7/2}}\frac{r^4}{l_s^7}\sqrt{\pi}\left(\frac{2r}{l_s}\right)^{-1} \textrm{Erf}\left(\frac{2r}{l_s}\right)+\mathcal{O}(\omega^6).
\end{align}
For odd $p$, we obtain 
\begin{align}
\tilde{V}_{c,1} (2r)
=&-\frac{\omega ^4}{8\pi} \frac{l_s^4}{r^2}
\left[1-\left(1+\left(\frac{2r}{l_s}\right)^{2}+\frac12\left(\frac{2r}{l_s}\right)^{4}\right)e^{-\left(2r/l_s\right)^2}\right]+\mathcal{O}(\omega^6),\\[2mm]
\tilde{V}_{c,3} (2r)
=&-\frac{\omega ^4}{16\pi^2}
\left[1-\left(1+\left(\frac{2r}{l_s}\right)^2\right)e^{-\left(2r/l_s\right)^2}\right]+\mathcal{O}(\omega^6),\\[2mm]
\tilde{V}_{c,5} (2r)
=&-\frac{\omega^4}{16\pi^3}\frac{r^2}{l_s^4}\left(1-e^{-\left(2r/l_s\right)^2}\right)+\mathcal{O}(\omega^6).
\end{align}
Note that the appearance of the factor $r^{-2}$ for $p=0,1$ does not imply the blow up of the effective potentials for small $r$. 
This is because our calculations so far are valid under the assumptions $r\omega, \omega l_s^2/r\ll1$, and therefore, $r$ cannot be extremely small.

\section{Shape of the effective potential}
\label{shape}

In the previous sections, we have  explicitly evaluated
 the integrals to obtain the effective potentials $\tilde{V}_p(2r)$ approximately for small $\omega$. 
 In this section, we investigate the shape of the effective potential $U_p(r)$, which includes the effect of the centrifugal force.
The formulas are valid as long as $\omega < r/l_s^2$ and $\omega r <1$ are satisfied. 
In the following, we focus our attention on the leading contribution for small $\omega$, that is, the contributions of order ${\cal O}(\omega^2)$. 
As long as $r$ is small compared to $l_s$, the SUGRA contribution $\tilde{V}_{c,p}(2r)$ is negligible. 
Note that for larger $r$, the SYM contributions $\tilde{V}_{o,p}(2r)$ decay exponentially, and the potential is dominated by the SUGRA contributions. 

We are interested in a possibility of two D$p$-banes to form
a stable (resonant) state  with a  length scale dynamically determined by the revolving motion. 
Our strategy to investigate such a possibility is the following. 
Analogous to the classical mechanics with the Newtonian potential, we consider 
\begin{equation}
U_p(r)\ :=\ \tilde{V}_p(2r)+V_{{\rm cent},p}(r), 
\end{equation}
where $V_{{\rm cent},p}(r)$ is the centrifugal potential for the D$p$-branes. 
If there is a local minimum for $U_p(r)$, there could be a non-trivial state which is stable, at least when the 
quantum tunneling effect is ignored. 
Since the lowest energy state consisting of two D$p$-branes is the static one which is half-BPS, the non-trivial state, if exists, should have 
positive energy. 

In order to study this, we need to take into account the effect of the centrifugal force. 
The centrifugal potential $V_{{\rm cent},p}(r)$ is derived from the SYM action with the revolving background $B_I$ for the adjoint scalar fields $\Phi_I$.
The background part of the SYM action in the Lorentzian signature is given by
\begin{equation}
-\frac1{2g_{\rm YM}^2}\int d^{p+1}x\ \textrm{Tr}(\partial_\mu B_I)^2+\cdots
\end{equation} 
where the revolving D$p$-brane is described by 
\begin{equation}
B_8=r\cos(\omega t)\cdot \sigma_3/l_s^2, \hspace{5mm} B_9=r\sin(\omega t)\cdot \sigma_3/l_s^2. 
\end{equation}
Substituting these backgrounds into the action, we obtain 
\begin{align}
V_{{\rm cent},p}(r,\omega)=\frac{r^2 \omega^2}{g_{\rm YM}^2l_s^4}=r^2\omega^2T_p. 
\end{align} 
This becomes a more familiar form when we fix the angular momentum per volume $L=2r^2\omega T_p$ and replace $\omega$ with $L/2r^2T_p$. 
To indicate the $g_s$-dependence explicitly, we denote $T_p=a_p/g_s$ where $a_p:=(2\pi)^{(1-p)/2}l_s^{-p-1}$. 
Accordingly, we define $L_0:=Lg_s$ which is independent of $g_s$. 
Then, the relation for the angular momentum can be rewritten as $L_0=2r^2\omega a_p$.
Replacing $\omega$ with $L_0/2r^2a_p$ in the $V_{{\rm cent},p}(r)$, we obtain 
\begin{align}
U_p (r)=\frac{L_0^2}{4a_pg_s r^2}+\tilde V_p(r)\Big|_{\omega=L_0/2r^2a_p}.
\end{align}
For large $r$, we know from the SUGRA analysis that $\tilde{V}_p(r)$ behaves as $-r^{p-7}$. 
This is much weaker than $V_{{\rm cent},p}(r)$ and $U_p(r)$ is repulsive.
Therefore resonant states, if exist, should 
be formed for small $r$ region where SYM contributions are important. 

The plots for various $p$ are given in Figure \ref{FigN1}.
In the figure, we set $T_p=1$ and $L=0.01$. 
We find that the potential $U_p(r)$ is always repulsive for $p\ge3$, even for small $r$. 
This implies that no resonant states are anticipated for $p \ge 3$. 
For $p=0,1$ and $2$, the situation is different and more interesting. 
For those cases, the potential changes its behavior and becomes attractive for small enough $r$. 
This may indicate the existence of a resonant state. 
Interestingly, such a state was discussed in \cite{Danielsson:1996uw, Kabat:1996cu}. 
The behavior of our potential seems to be consistent with their result.

\subsection{Multiple D$p$-branes}
\begin{figure}[htpb]
\center
\begin{tabular}{cc}
\begin{minipage}{.4\hsize}
\center
\includegraphics [scale=.5]
 {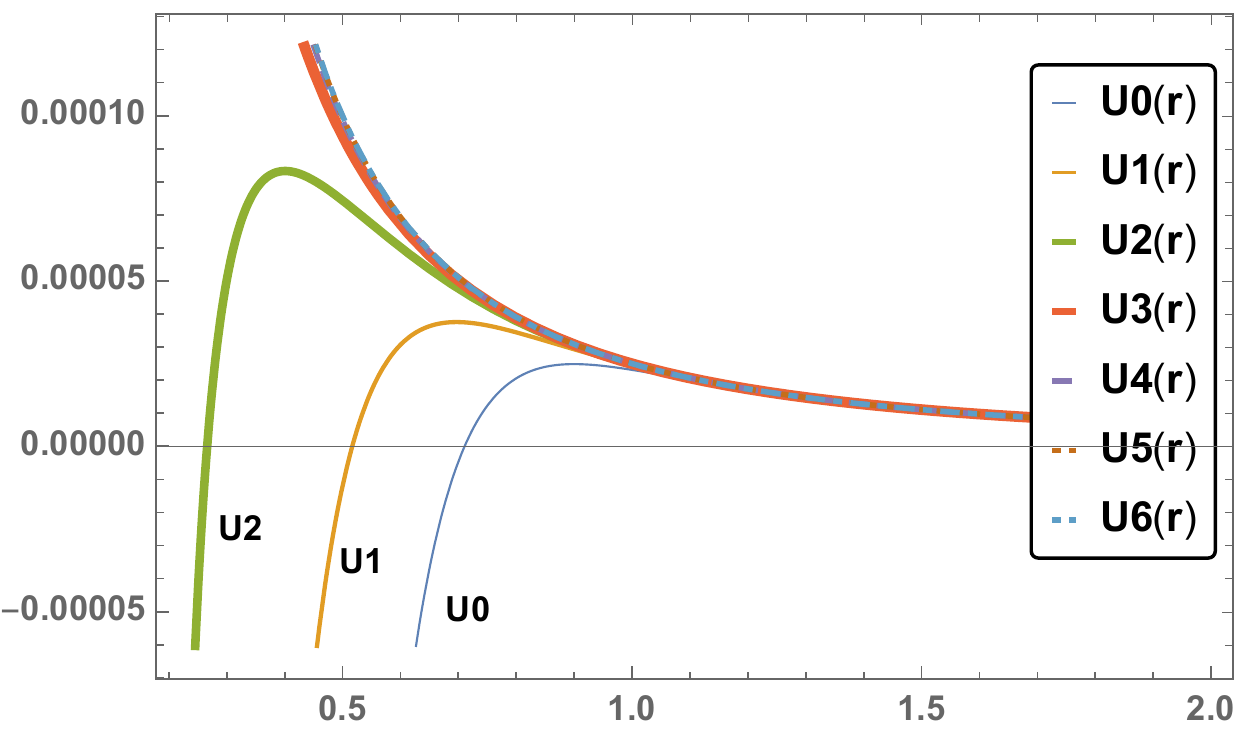}
\caption{
The potential $U_p(r)$ with $T_p=1$ and $L=0.01$.
}
\label{FigN1}
\end{minipage}
\hspace{1cm}
\begin{minipage}{.4\hsize}
\center
\includegraphics [scale=.5]
 {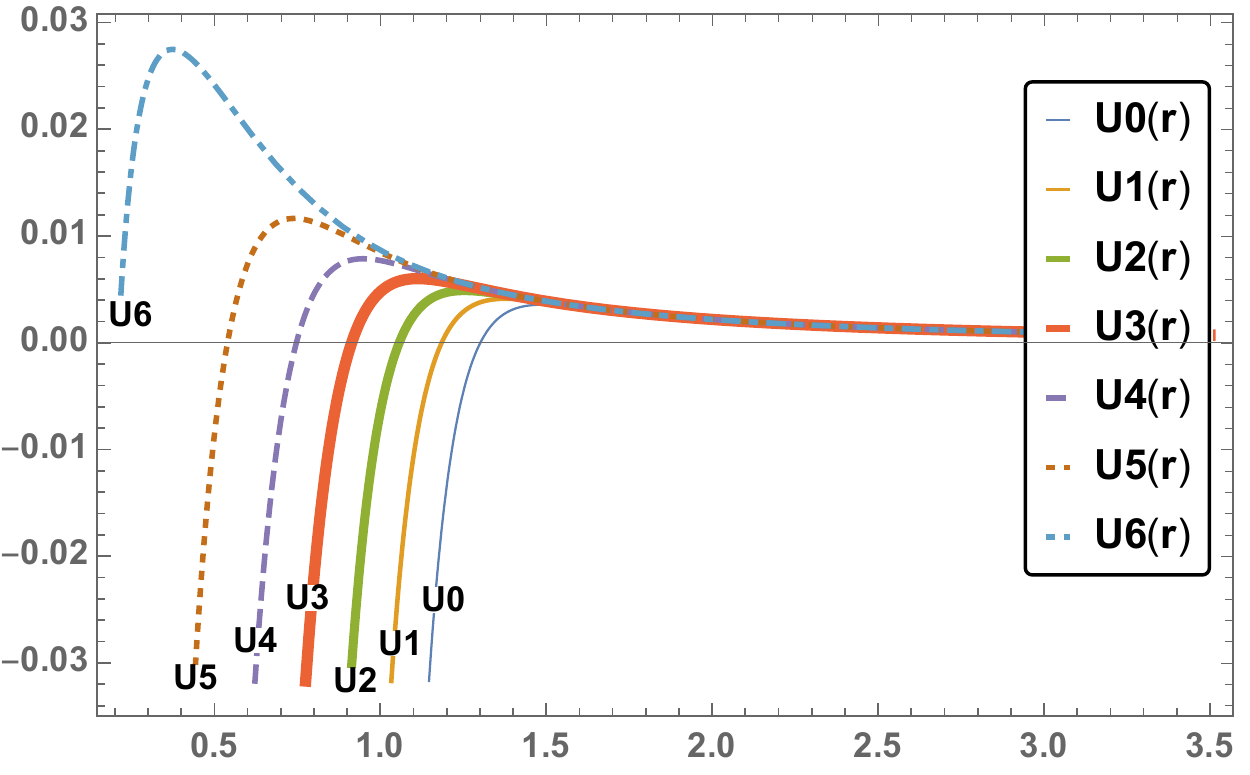}
\caption{
The potential $U_{N,p}(r)$ with $T_p=1$, $L=0.01$ and $N=350$.
}
\label{FigN350}
\end{minipage}
\end{tabular}
\end{figure}

We can generalize the previous results of $U_p(r)$ to $U_{N,p} (r)$, where
a pair of stacks of $N$ D$p$-branes is considered
 instead of  a pair of just a single D$p$-brane. 
Then the number of open strings stretched between the pair becomes $N^2$ and thus $\tilde V_p(2r)$ is  multiplied by $N^2$, while the centrifugal potential $V_{{\rm cent},p}(r)$ is simply multiplied by $N$. 
As a result, the potential becomes 
\begin{align}
U_{N,p} (r):=\frac{N L_0^2}{4a_pg_s r^2}+N^2\tilde V_p(2r).
\end{align}
The shapes of $U_{N,p}(r)$ are depicted in figure \ref{FigN350}. 
%
Here also we set $T_p=1$, $L=0.01$ and $N=350$.
We can see that, for a sufficiently large $N$, the potential becomes negative 
and falls off for smaller $r$. 

To see the behavior of $U_{N,p}(r)$ for small $r$ in more detail, we expand $U_{N,p}(r)$ with respect to $r$. 
Since $\tilde{V}_p(2r)$ was determined by expanding in terms of $\omega$, 
the further $r$-expansion is valid as long as $\omega\ll r/l_s^2$ is satisfied. 

Recall that the leading order of $\tilde{V}_{p}(2r)$ is given by
\begin{eqnarray}
\tilde{V}_{p}(2r) 
&=& -\frac4\pi(4\pi)^{-\frac p2}\frac{\sqrt{\pi}}{\Gamma(\frac{p+1}2)}\omega^2l_s^{1-p}\left( \frac{2r}{l_s} \right)^2f_{p,1}\left( \frac{2r}{l_s} \right)+{\cal O}(\omega^4), 
\end{eqnarray}
where the analytic continuation of $\omega$ have performed.
At small $x$, the function $f_{p,1}(x)$ behaves as
\begin{equation}
f_{p,1}(x)\ \sim\ \left\{
\begin{array}{lc}
\displaystyle{\frac{\pi}{2x}}-\sqrt{\pi}, & (p=0) \\ [3mm]
\displaystyle{\frac12\log \frac{e^{-\gamma}}x}, & (p=1) \\ [3mm]
\displaystyle{\frac12\Gamma({\textstyle\frac{p-1}{2}})}. & (p\ge2)
\end{array}
\right. 
\end{equation}
For the details, see Appendix \ref{app-small r}. 
Thus, for $p \ge 2$, we have $V_p(2r) \propto \omega^2 r^2$. For $p=0, 1$, it has a different $r$-dependence;
$V_0(2r) \propto \omega^2 r$ and $V_1(2r)  \propto \omega^2 r^2 \log r.$

\vspace{5mm}

Now we discuss  competition between $\tilde{V}_p(2r)$ and $V_{{\rm cent},p}(r)$. 
Let us first consider $p\ge2$ cases.
We have $N^2\tilde{V}_p(2r)\sim-c_pN^2\omega^2r^2$ for small $r$, where 
\begin{equation}
c_p\ :=\ \frac{32}{p-1}(4\pi)^{-(p+1)/2}l_s^{-p-1}. 
\end{equation}
For a fixed angular momentum, the potential $\tilde{V}_p(2r)$ becomes 
\begin{equation}
N^2\tilde{V}_p(2r)\ \sim\ -\frac{c_pN^2}{a_p}\frac{L_0^2}{4a_pr^2}. 
\end{equation}
Therefore, the potential $U_{N,p}(r)$ is 
\begin{equation}
U_{N,p}(r)\ \sim\ \left( \frac1{g_s}-\frac{c_pN}{a_p} \right)\frac{NL_0^2}{4a_pr^2}. 
\end{equation}
If this is negative, then the attractive interaction wins for small $r$. 
We find that the 't~Hooft coupling $g_sN$ governs the sign of the potential for small $r$. 
Indeed, if the following condition is satisfied, 
\begin{equation}
g_sN\ >\ \frac{a_p}{c_p}\ =\ 2^{-(7-p)/2}(p-1)\pi
\end{equation}
the potential becomes negative for small $r$. 
The right-hand side takes the smallest value $2^{-5/2}\pi=0.555$ for $p=2$. 
Therefore, we need an ${\cal O}(1)$ value for $g_sN$ if we would like to have an attractive force for small $r$, which
is a necessary condition for  a resonant state to be formed. 
The requirement $g_sN={\cal O}(1)$ suggests that the backreaction of the D-brane system to the space-time metric may not be neglected. 
Indeed, since the Newtonian potential created by $N$ D$p$-branes is roughly given by  the form (assuming that the dimension is adjusted by $l_s$)
\begin{equation}
\frac{G_{\rm N}T_p}{r^{7-p}}\ \sim\ \frac{g_sN}{r^{7-p}}, 
\end{equation}
a typical scale of the curvature of the space-time is given by
 $(g_sN)^{1/(7-p)}l_s$. 
Thus in order to discuss the existence of resonant states in a controlled manner, it will be necessary to consider curved background space-time in which D-branes are revolving. 

Next, let us consider a case of $p=0$. 
The potential becomes 
\begin{equation}
U_0\ \sim\
 \left( \frac{1}{g_s} + \frac{16N}{\sqrt{2}\pi} - \frac{4N}{\sqrt{2\pi}r/l_s} \right)
 \frac{NL_0^2}{4a_0r^2}. 
\end{equation}
Remarkably, the potential becomes attractive around $r/l_s\sim g_sN$ for any value of $g_sN$. 
Therefore, by taking $g_sN$ to be small, our analysis can be justified. 

Finally, the potential for $p=1$ turns out to be 
\begin{equation}
U_1\ \sim\
 \left( \frac{1}{g_s} - \frac{4N}{\pi}\log\frac{e^{-\gamma}l_s}{2r} \right)\frac{NL_0^2}{4a_1r^2}. 
\end{equation}
This becomes attractive for a region of $r$ smaller than 
\begin{equation}
r_c\ \sim\ \frac12\exp\left(-\frac{\pi}{4g_sN}-\gamma \right)l_s. 
\end{equation}
It is curious that the right-hand side has a non-perturbative form for $g_sN$. 

It is interesting that the potential becomes always negative for $p=0$ and $p=1$
for small $r$, which indicates the existence of resonant states of two revolving D$p$-branes. 
\section{Conclusions and Discussions}
\label{conclusion}

In this paper, we have obtained the effective potential of revolving D$p$-branes.
The interactions between D-branes can be evaluated either by closed string exchange for long distance $r>l_s$
or by open string 1-loop amplitude for short distance $r<l_s$, and they are related by a modular transformation.
In order to evaluate the effective potential interpolating these two regions, we used the technique of 
{\it partial} modular transformation to sum up both of open and closed string contributions without a double counting. 
The shape of the effective potential $U_{N, p}(r)$, which includes the effect of the centrifugal force, 
is drawn for small angular frequency $\omega$  in a nonrelativistic region. 
The interaction is repulsive for large $r$, but we find that it can become attractive for small $r$.

The shapes  are different between $p=0, 1$ and $p \ge 2$.
Since the infrared effects are stronger for $p=0, 1$,
the potential falls down faster around $r \sim 0$.
Especially for $p=1$, there is an additional logarithmic factor $\log(r)$
in the effective potential and the typical distance $r=r_c$
where the interaction changes from repulsive ($r>r_c$) to attractive
($r<r_c$) is given in a nonperturbative form
$r_c \propto \exp(-\pi/4g_s N) l_s$.
On the other hand, for $p \ge 2$, 
the typical shape of the
potential is given by $-c/r^2$ for small $r$.
Thus it is classically unstable. 
Quantum mechanically, depending on the numerical coefficient $c >0$, the system is either unstable or
has a bound state \cite{Landau:1991wop}. In our system the potential is more complicated, since it
becomes positive in an intermediate region of $r$, and
we need more detailed analysis to see whether it has a metastable bound state or not. 
Furthermore, in the paper, we 
have evaluated the integral in $U_{N,p}(r)$ for small $r$ by using $\omega/r$ expansions.
Thus angular frequency must be small, $\omega <r/l_s^2$, and the approximation is not valid 
for $r < l_s^2 \omega$. 

In the following discussion, we look at the behavior of the potential at $r \sim 0$
by using another expansion with respect to $r/\omega$. 
First we can easily see that the effective potential vanishes at $r=0$; $\tilde{V}_{o}(0)= \tilde{V}_{B,p}(0)+\tilde{V}_{F,p}(0) =0$
by shifting the integration variable, $k_\tau \rightarrow k_\tau \pm \omega$.
Thus, the unstable behavior in the $\omega/r$ expansion may be merely ostensible. 
In order to further understand the behavior at $r \sim 0$,
 let us go back to the calculation of the bosonic part, eq. (\ref{SYM-boson}).
It consists of integrals 
\begin{equation}
-\int_{\Lambda^{-2}}^\infty\frac{dt}{t}\int\frac{d^{p+1}k}{(2\pi)^{p+1}}e^{-tf(k,\theta)}
\label{int-small k}
\end{equation}
with a singular function at $(k, r) \sim (0,0)$:
\begin{equation}
f(k,\theta)\ :=\ k^2+4r^2+\omega^2-\frac{8(r\omega)^2}{k^2+4r^2}-\sqrt{4\omega^2\cos^2\theta\, k^2+\left( \frac{8(r\omega)^2}{k^2+4r^2} \right)^2},
\end{equation}
where we set $k_\tau=k \cos\theta$. 
Around $k=0$, this behaves as 
\begin{equation}
f(k)\ =\ 4r^2-3\omega^2+\left( 1-\cos^2\theta+\frac{\omega^2}{r^2} \right)k^2+{\cal O}(k^4). 
\end{equation}
Note that $f(k)$ becomes negative when $\omega/r>2/\sqrt{3}$,
 which implies that the $t$-integral diverges. 
To illustrate this, suppose that $\omega/r$ is large. 
Then, the above integral can be estimated for large $t$ as 
\begin{eqnarray}
& & -\int^\infty\frac{dt}{t}\int\frac{d^{p+1}k}{(2\pi)^{p+1}}e^{-t(f(0)+(1-\cos^2\theta+(\omega/r)^2)k^2)} \nonumber \\ [2mm] 
&\sim& -(4\pi)^{(p+1)/2}\left( \frac r\omega \right)^{p+1}\int^\infty\frac{dt}{t}t^{-(p+1)/2}e^{-tf(0)}. 
\end{eqnarray}
Indeed, this integral diverges when $f(0)=4r^2-3\omega^2<0$. 
Recall that, if we fix the angular momentum $L$, then $\omega$ grows as $r^{-2}$. 
Therefore, the potential $U_{N,p}(r)$ has a problem for too small $r$ with fixed $L$.

To see what happens at $\omega/r=2/\sqrt{3}$, consider the function 
\begin{equation}
\int_1^\infty\frac{dt}{t}t^{-(p+1)/2}e^{-zt}\ =\ E_{(p+3)/2}(z) 
\end{equation}
of a complex variable $z$. 
The integral in the left-hand side is well-defined for ${\rm Re}(z)>0$. 
We find that $z=0$ is a branch point. 
The non-analytic part turns out to be 
\begin{equation}
E_{(p+3)/2}(z)\ \sim\ \left\{
\begin{array}{lc}
\displaystyle{-\frac{(-z)^{n+1}}{(n+1)!}\log z}, & (p=2n+1) \\ [4mm]
\displaystyle{(-1)^{n+1}\frac{\pi}{\Gamma(n+\frac32)}z^{n+\frac12}.} & (p=2n)
\end{array}
\right. 
\end{equation}
Therefore, if we define $U_{N,p}(r)$ for small $r$ via the analytic continuation, then it takes complex value for $\omega/r>2/\sqrt{3}$. 
This may be a signal of an instability of the system. 
Since this behavior comes from the fact that the integral (\ref{int-small k}) is not well-defined for small momentum $k$, it is tempting to speculate that the shape of the D-branes might be deformed due to the revolving motion. 
Finally we note that this kind of instability does not occur for the fermionic part of the effective action $\tilde{V}_{F,p}(2r)$.
It is specific to the bosonic part in which the kinetic terms of the gauge bosons and scalars  are mixed due to the revolving motion.
This situation is similar to \cite{Hashimoto:2003xz}. 
The effective potential vanishes just at $r=0$, but the instability itself seems to be present. 
We want to come back to the origin of the instability in future investigations. 

\vspace{1cm}

{\Large \bf Acknowledgements}

\vspace{3mm}

This work is supported in part by Grants-in-Aid for Scientific Research No. 16K05329, No. 18H03708 and No. 19K03851 from the Japan Society for the Promotion of Science.


\appendix

\section{Various functions for SYM contributions}
\label{app-SYM}

In subsection \ref{sec:SYM}, we need to determine the explicit form of functions defined as 
\begin{equation}
f_{p,m}(x)\ :=\ x^{1+p-2m}\int_0^\infty dk\,e^{-x^2(k^2+1)}\frac{k^p}{(k^2+1)^m}. 
   \label{f_{p,m}(x)-App}
\end{equation}
We can rewrite them as 
\begin{eqnarray}
f_{p,m}(x) 
&=& x^{1+p-2m}\int_0^\infty dk\,e^{-x^2(k^2+1)}\frac{k^{p-2}(k^2+1)-k^{p-2}}{(k^2+1)^m} \nonumber \\
&=& f_{p-2,m-1}(x)-x^2f_{p-2,m}(x). 
\label{recurrence-SYM}
\end{eqnarray}
These recurrence relations allow us to determine $f_{p,m}(x)$ from $f_{0,m}(x)$ or $f_{1,m}(x)$ according to whether $p$ is even or odd, respectively. 
Note that, to obtain $f_{p,m}(x)$ for $p\ge2$ and $-1\le m\le 4$, we need $f_{0,m}(x)$ or $f_{1,m}(x)$ with $m<-1$. 

\subsection{$f_{0,m}(x)$}
For $m=0$, this is just a Gaussian integral,
\begin{align}
f_{0,0}(x)=\frac{\sqrt \pi}{2}e^{-x^2}.
\end{align}
The functions $f_{0,m}(x)$ satisfy
\begin{align}
f_{0,m-1}(x)\ =\ -\frac12x^{2-2m}\frac{d}{dx}\left( x^{2m-1}f_m(x) \right). 
   \label{f_{0,m}(x)}
\end{align}
From this relation, we obtain $f_{0,m}(x)$ for $m<0$ easily from $f_{0,0}(x)$. 
For example, 
\begin{eqnarray}
f_{0,-1}(x) &=& \frac{\sqrt{\pi}}{4}(1+2x^2)e^{-x^2}, \\
f_{0,-2}(x) &=& \frac{\sqrt{\pi}}{8}(3+4x^2+4x^4)e^{-x^2}, \\
f_{0,-3}(x) &=& \frac{\sqrt{\pi}}{16}(15+18x^2+12x^4+8x^6)e^{-x^2}, \\
f_{0,-4}(x) &=& \frac{\sqrt{\pi}}{32}(105+120x^2+72x^4+32x^6+16x^8)e^{-x^2}. 
\end{eqnarray}

Next, we consider $m>0$. 
By integrating (\ref{f_{0,m}(x)}), we obtain 
\begin{equation}
f_{0,1}(x)\ =\ \frac{\pi}{2x}(1-{\rm Erf}(x)), 
\end{equation}
where 
\begin{align}
\textrm{Erf}(x):=\frac{2}{\sqrt{\pi}}\int_0^x dt\, e^{-t^2}, 
\end{align}
and the integration constant is specified by $\displaystyle{\lim_{x\to \infty} f_m(x)=0}$ which is obvious from the original integral form (\ref{f_{p,m}(x)-App}).
The functions with $m\ge2$ can be obtained similarly. 
It is also convenient to use the following recurrence relations 
\begin{equation}
f_{0,m}(x)\ =\ \frac1{x^2}\left( 1-\frac{1+2x^2}{2(m-1)} \right)f_{0,m-1}(x)+\frac1{(m-1)x^2}f_{0,m-2}(x). 
\end{equation}
This can be derived as follows;
\begin{eqnarray}
f_{0,m}(x) 
&=& x^{1-2m}\int_0^\infty dk\,e^{-x^2(k^2+1)}\frac{(k^2+1)-k^2}{(k^2+1)^m} \nonumber \\
&=& \frac1{x^2}f_{0,m-1}(x)-x^{1-2m}\int_0^\infty dk\,e^{-x^2(k^2+1)}\frac{k^2}{(k^2+1)^m}, 
\end{eqnarray}
where the last integral can be calculated as 
\begin{eqnarray}
& & -x^{1-2m}\int_0^\infty dk\,e^{-x^2(k^2+1)}\frac{k^2}{(k^2+1)^m} \nonumber \\ [2mm] 
&=& -\frac{x^{1-2m}}{2(m-1)}\int_0^\infty dk\,e^{-x^2(k^2+1)}\frac{1-2x^2k^2}{(k^2+1)^{m-1}} \nonumber \\
&=& -\frac1{2(m-1)x^2}f_{0,m-1}(x)+\frac1{(m-1)x^2}f_{0,m-2}(x)-\frac1{m-1}f_{0,m-1}(x). 
\end{eqnarray}

We can obtain
\begin{align}
f_{0,2}(x)=&\frac{\sqrt \pi}{2x^2} e^{-x^2}+\frac{\pi}{4x^3} (1-2x^2)\left(1-\textrm{Erf}(x)\right) \\
f_{0,3}(x)=&\frac{\sqrt \pi}{8x^4} e^{-x^2}(3-2x^2)+\frac{\pi}{16x^5} (3-4x^2+4x^4) \left(1-\textrm{Erf}(x)\right) \\
f_{0,4}(x)=&\frac{\sqrt \pi}{48x^6} e^{-x^2}(15-8x^2+4x^4)+\frac{\pi}{96x^7} (15-18x^2+12x^4-8x^6) \left(1-\textrm{Erf}(x)\right). 
\end{align}

\subsection{$f_{1,m}(x)$}
This can be written as 
\begin{align}
f_{1,m}(x)\ =\ x^{2-2m}\int_0^\infty dk\,e^{-x^2(k^2+1)}\frac{k}{(k^2+1)^{m}}
=&\ \frac12 x^{2-2m}E_m \left(x^2\right),
\end{align}
where 
\begin{equation}
E_m(x):=\int_1^\infty dt\, t^{-m} e^{-xt}.
\end{equation}
For $m\geq1$, $E_{m+1}(x)$ satisfy
\begin{equation}
E_{m+1}(x)\ =\ -\frac xmE_m(x)+\frac{1}{m}e^{-x}. 
\end{equation}
This recurrence relation can be solved in terms of $E_1(x)$. 
The solution is 
\begin{align}
E_{m+1}(x)
=\frac{(-x)^m}{m!}E_1(x)-\frac{e^{-x}}{x}\sum_{l=1}^{m}\frac{(m-l)!}{m!}(-x)^{l}.
\end{align}
By using them, we obtain 
\begin{eqnarray}
f_{1,1}(x) 
&=& \frac12E_1(x^2), \\
f_{1,2}(x) 
&=& \frac1{2x^2}e^{-x^2}-\frac1{2}E_1(x^2), \\
f_{1,3}(x) 
&=& \frac{1-x^2}{4x^4}e^{-x^2}+\frac1{4}E_1(x^2), \\
f_{1,4}(x) 
&=& \frac{2-x^2+x^4}{12x^6}-\frac1{12}E_1(x^2). 
\end{eqnarray}

For $m\le0$, we use 
\begin{equation}
E_0(x)\ =\ \frac1xe^{-x}, \hspace{1cm} E_{m-1}(x)\ =\ -\frac d{dx}E_m(x). 
\end{equation}
From them, we obtain 
\begin{eqnarray}
f_{1,0}(x) 
&=& \frac12e^{-x^2}, \\
f_{1,-1}(x) 
&=& \frac12(1+x^2)e^{-x^2}, \\
f_{1,-2}(x) 
&=& \frac12(2+2x^2+x^4)e^{-x^2}, \\
f_{1,-3}(x) 
&=& \frac12(6+6x^2+3x^4+x^6)e^{-x^2}, \\
f_{1,-4}(x) 
&=& \frac12(24+24x^2+12x^4+4x^6+x^8)e^{-x^2}. 
\end{eqnarray}

\section{Various functions for SUGRA contributions}
\label{app-SUGRA}

For the calculations of the SUGRA contributions $\tilde{V}_c(2r)$, we need to determine 
\begin{equation}
g_{\alpha}(x)\ :=\ \int_1^\infty ds\,s^{-\alpha}e^{-x^2/s}, 
\end{equation}
where $\alpha:=(9-p)/2$. 
The recurrence relations 
\begin{eqnarray}
g_\alpha(x) 
&=& \int_0^1dt\,t^{\alpha-2}e^{-x^2t} \nonumber \\
&=& -\frac1{x^2}t^{\alpha-1}e^{-x^2t}\Big|_0^1+\frac{\alpha-2}{x^2}\int_0^1dt\,t^{\alpha-3}e^{-x^2t} \nonumber \\
&=& -\frac1{x^2}e^{-x^2}+\frac{\alpha-2}{x^2}g_{\alpha-1}(x)
   \label{recurrence-SUGRA}
\end{eqnarray}
for $\alpha>2$ imply that we only need to determine $g_{3/2}(x)$ or $g_2(x)$, depending on whether $p$ is even or odd, respectively. 

\subsection{$g_{3/2}(x)$}

This can be rewritten as 
\begin{align}
g_{3/2}(x)
=&\int^\infty_1 ds \, s^{-3/2}e^{-x^2/s}
=\frac{\sqrt \pi}x \, \textrm{Erf}(x).
\end{align}
Then, the solutions of the recurrence relations (\ref{recurrence-SUGRA}) is 
\begin{align}
g_{m+3/2}(x)
=&\frac{\Gamma(m+1/2)}{x^{2m+1}}\textrm{Erf}(x)-\frac{e^{-x^2}}{x^2}\sum_{l=0}^{m-1}\frac{\Gamma(m+1/2)}{\Gamma(m-l+1/2)}x^{-2l}.
\end{align}
Explicitly, 
\begin{eqnarray}
g_{5/2}(x) 
&=& \frac{\sqrt{\pi}}{2x^3}{\rm Erf}(x)-\frac1{x^2}e^{-x^2}, \\
g_{7/2}(x) 
&=& \frac{3\sqrt{\pi}}{4x^5}{\rm Erf}(x)-\frac{3+2x^2}{2x^4}e^{-x^2}, \\
g_{9/2}(x) 
&=& \frac{15\sqrt{\pi}}{8x^7}{\rm Erf}(x)+\frac{15+10x^2+4x^4}{4x^6}e^{-x^2}. 
\end{eqnarray}

\subsection{$g_2(x)$}
This is simply 
\begin{align}
g_2(x)=\frac{1-e^{-x^2}}{x^2}.
\end{align}
Thus, for $m\geq1$, $g_{m+2}(x)$ is given as
\begin{align}
g_{m+2}(x)=\frac{m!}{x^{2m+2}}-\frac{e^{-x^2}}{x^2}\sum_{l=0}^{m}\frac{m!}{(m-l)!}x^{-2l}.
\end{align}
Explicitly, 
\begin{eqnarray}
g_3(x) 
&=& \frac1{x^4}-\frac{1+x^2}{x^4}e^{-x^2}, \\
g_4(x) 
&=& \frac2{x^6}-\frac{2+2x^2+x^4}{x^6}e^{-x^2}. 
\end{eqnarray}

\section{${\cal O}(\omega^2)$ contributions for small $r$}
\label{app-small r}

The effective potential $\tilde{V}_p(2r)$ at the leading order in $\omega$ is given by $f_{p,1}(x)$. 
They can be determined recursively, as explained in Appendix \ref{app-SYM}. 
When we are only interested in the leading order terms in $f_{p,1}(x)$ with respect to $x$, they can be derived more easily as follows. 

It is easy to obtain 
\begin{equation}
f_{p,0}(x)\ =\ \frac12\Gamma({\textstyle\frac{p+1}2})e^{-x^2}. 
\end{equation}
Then, the recurrence relation (\ref{recurrence-SYM}) becomes 
\begin{equation}
f_{p,1}(x)\ =\ \frac12\Gamma({\textstyle \frac{p-1}2})e^{-x^2}-x^2f_{p-2,1}(x). 
\end{equation}

First, we assume that $p$ is even. 
We found in Appendix \ref{app-SYM} that 
\begin{eqnarray}
f_{0,1}(x) 
&=& \frac\pi{2x}(1-{\rm Erf}(x))\ =\ \frac\pi{2x}-\sqrt{\pi}+{\cal O}(x^2), \\ [2mm]
f_{2,1}(x) 
&=& \frac{\sqrt{\pi}}2e^{-x^2}-x^2f_{0,1}(x)\ =\ \frac{\sqrt{\pi}}{2}+{\cal O}(x). 
\end{eqnarray}
Recursively, we can show that $f_{p,1}(x)={\cal O}(1)$ for $p\ge2$. 
Note that the qualitative behavior is different only for $p=0$. 

For odd $p$, 
\begin{eqnarray}
f_{1,1}(x) 
&=& \frac12E_1(x^2)\ =\ -\frac12(\gamma+\log x)+{\cal O}(x), \\ [2mm]
f_{3,1}(x) 
&=& \frac12e^{-x^2}-x^2f_{1,1}(x)\ =\ \frac12+{\cal O}(x^2\log x). 
\end{eqnarray}
Recursively, we can show that $f_{p,1}(x)={\cal O}(1)$ for $p\ge3$. 
We find that there exists a logarithmic correction to the leading behavior for $p=1$. 
In summary, the leading behavior of $f_{p,1}(x)$ is 
\begin{equation}
f_{p,1}(x)\ \sim\ \left\{
\begin{array}{lc}
\displaystyle{\frac\pi{2x}-\sqrt{\pi}}, & (p=0) \\ [3mm] 
\displaystyle{\frac12\log \frac{e^{-\gamma}}x}, & (p=1) \\ [3mm] 
\displaystyle{\frac12\Gamma({\textstyle \frac{p-1}2})}. & (p\ge2)
\end{array}
\right. 
\end{equation}

\end{document}